\documentclass[twocolumn,conference]{IEEEtran}
\PassOptionsToPackage{ruled}{algorithm2e}
\usepackage[T1]{fontenc}
\usepackage[latin9]{inputenc}
\usepackage{algorithm2e}
\usepackage{amsmath}
\usepackage{amsthm}
\usepackage{amssymb}
\usepackage{graphicx}
\usepackage{epstopdf}

\makeatletter

\providecommand{\tabularnewline}{\\}

\theoremstyle{plain}
\newtheorem{thm}{\protect\theoremname}
\theoremstyle{plain}
\newtheorem{prop}[thm]{\protect\propositionname}

\usepackage{cite}
\usepackage{siunitx}

\DeclareMathOperator{\maximize}{maximize}

\DeclareMathOperator{\st}{subject \, to}

\DeclareMathOperator{\diag}{diag}
\DeclareMathOperator{\Var}{\mathbb{V}ar}

\newcommand{\herm}{^{{\dagger}}}
\newcommand{\trans}{^{\mbox{\scriptsize T}}}

\makeatother

\providecommand{\propositionname}{Proposition}
\providecommand{\theoremname}{Theorem}

\begin{document}
\title{Power Control for Multigroup Multicast Cell-Free Massive MIMO Downlink}
\author{\IEEEauthorblockN{Muhammad~Farooq\IEEEauthorrefmark{1}, Markku Juntti\IEEEauthorrefmark{2}
and Le-Nam~Tran\IEEEauthorrefmark{1}}\IEEEauthorblockA{\IEEEauthorrefmark{1}School of Electrical and Electronic Engineering,
University College Dublin, Ireland\\
Email: muhammad.farooq@ucdconnect.ie, nam.tran@ucd.ie}\IEEEauthorblockA{\IEEEauthorrefmark{2}Centre for Wireless Communications, University
of Oulu, P.O.Box 4500, FI-90014 University of Oulu, Finland\\
Email: markku.juntti@oulu.fi}}
\IEEEspecialpapernotice{Invited Paper}
\maketitle
\begin{abstract}
We consider a multigroup multicast cell-free multiple-input multiple-output
(MIMO) downlink system with short-term power constraints. In particular,
the \emph{normalized conjugate beamforming} scheme is adopted at each
access point (AP) to keep the downlink power strictly under the power
budget regardless of small scale fading. In the considered scenario,
APs multicast signals to multiple groups of users whereby users in
the same group receive the same message. Under this setup, we are
interested in maximizing the minimum achievable rate of all groups,
commonly known as the max-min fairness problem, which has not been
studied before in this context. To solve the considered problem, we
first present a bisection method which in fact has been widely used
in previous studies for cell-free massive MIMO, and then propose an
accelerated projected gradient (APG) method. We show that the proposed
APG method outperforms the bisection method requiring lesser run time
while still achieving the same objective value. Moreover, the considered
power control scheme provides significantly improved performance and
more fairness among the users compared to the equal power allocation
scheme.
\end{abstract}


\section{Introduction}

Cell-free massive multiple-input multiple-output (MIMO), where multiple
users are simultaneously served by a larger number of access points
(APs) in the same time spectrum resource, was first introduced in
\cite{Ngo2017} and is being considered a promising technique for
beyond 5G networks. In principle, cell-free massive MIMO incorporates
the inherent advantages of both network MIMO and colocated massive
MIMO \cite{Mazretta2010}, and therefore can achieve high coverage
area, spectral efficiency (SE) and energy efficiency \cite{Ngo2018EE}.
In particular, cell-free massive MIMO has the capability to provide
users with nearly uniform service. Despite several benefits, scalability
remains a challenge in cell-free massive MIMO since (i) the increasing
number of high-capacity backhaul links are required to connect APs
to the central processing unit (CPU), and (ii) large-scale resource
allocation problems need to be solved at the CPU to deliver the best
performance. The latter issue makes the fundamental performance of
cell-free massive MIMO limited to only the small-scale systems \cite{Bashar2018}.

In many practical situations, a group of users may be interested in
the same information like headline news, weather update, live streaming,
financial data, etc., which has motivated the study of multigroup
multicast systems in massive MIMO \cite{Doan2017,Sadeghi2018MGMC}.
For cell-free massive MIMO, the first noticeable work in multigroup
multicasting was carried out in \cite{Doan2017}, where a closed-from
expression of the achievable rate for single-antenna APs and single-antenna
users was derived. Moreover, the \emph{normalized conjugate beamforming}
scheme \cite{Interdonato2016stpc} was used in \cite{Doan2017}, which
is devised on the basis of short-term power constraint (STPC). Note
that the goal of the STPC policy is to ensure that the transmit power
is always under the maximum budget regardless of instant channel gain
and thus is of more practical importance \cite{Khoshnevisan2011}.
This is in opposite to beamformers derived from a long-term power
constraint (LTPC) policy which has been adopted in many previous studies
\cite{Ngo2017,Ngo2018EE}. It was shown in \cite{Interdonato2016stpc}
that normalized conjugate beamforming outmatches the common conjugate
beamforming when the number of APs is moderate as it hardens the effective
channel gains at users.

Power control for multigroup multicast cell-free massive MIMO systems
has not been studied. Doan \emph{et al.} in \cite{Doan2017} derived
the achievable rate based on the assumption that the downlink power
is equally allocated to all groups, which is often termed as \emph{equal
power allocation} (EPA). In \cite{Sadeghi2018MGMC}, Sadeghi \emph{et
al.} designed the precoders to maximize the minimum SE which is commonly
known as the \emph{max-min fairness} problem. To the best of our knowledge,
no prior literature has discussed the power control for max-min fairness
in multigroup multicasting cell-free massive MIMO which is our problem
of interest in this paper.

In this paper, we consider a multigroup multicast cell-free MIMO downlink
system using time division duplexing (TDD). Users in a group send
\emph{the same pilot sequence} to APs in the uplink for channel estimation
purpose. Based on the channel estimates, APs will form different beams
to different groups. In this considered system model, we derive an
achievable rate in closed form and formulate the power control problem
for max-min fairness based on STPC policy. To solve this problem,
we first present a bisection method which is popular in the context
of power control for cell-free massive MIMO \cite{Ngo2017,Ngo2018EE}.
However, such a method is only suitable for cell-free massive MIMO
of moderate sizes. To overcome this issue, we then propose a low-complexity
algorithm based on the accelerated projected gradient (APG) framework
\cite{Li2015,Farooq2020APG}. Simulation results demonstrate that
the proposed power control algorithm can offer significant performance
improvements over the EPA scheme in the considered scenarios. 

\emph{Notation}s: Standard notations are used in this paper. Bold
lower and upper case letters represent vectors and matrices. $\mathcal{CN}(\mathbf{0},\mathbf{R})$
denotes the multivariate circularly symmetric complex Gaussian random
distribution with zero mean and covariance matrix $\mathbf{R}$. $\mathbb{R}^{x\times y}$
represents the space of real matrices with the dimensions $x\times y$.
$\mathbf{X}^{\ast}$, $\mathbf{X}\trans$ and $\mathbf{X}\herm$ stand
for the conjugate, transpose and conjugate transpose (Hermitian) of
$\mathbf{X}$, respectively. The ``$+$'' sign in the subscript
of a space implies that all elements of that space are positive. $\mathbb{E}\{X\}$
denotes the expectation or mean of random variable $X$. $x_{i}$
is the $i$-th entry of vector $\mathbf{x}$; $[\mathbf{X}]_{i,j}$
is the entry at the $i$-th row and $j$-th column of $\mathbf{X}$.
$\Vert\cdot\Vert$ represents the Euclidean norm; $|\cdot|$ is the
absolute value of the argument. The operator $\diag$ converts a vector
intro a diagonal matrix. $\mathbf{I}_{N}$ is the $N\times N$ identity
matrix. $\ln(\cdot)$ denotes the natural logarithm.

\section{System Model and Problem Formulation}

\subsection{System Model}

Consider a cell-free massive MIMO scenario where $M$ single-antenna
APs are connected to a CPU via a perfect back-haul link. The APs coherently
transmit $N$ independent messages to $N$ groups of users in a TDD
mode. Note that all users in the same group receive the same message.
The number of single-antenna users in the $n$-th group is denoted
by $K_{n}$. Throughout the paper, we note that the notation $n_{k}$
refers to the $k$-th user in the $n$-th group. In this regard, the
channel coefficient between the $m$-th AP and the $k$-th user in
the $n$-th group is modeled as 
\begin{equation}
h_{mn_{k}}=\zeta_{mn_{k}}^{1/2}g_{mn_{k}},\label{eq:channel}
\end{equation}
where $\zeta_{mn_{k}}$ and $g_{mn_{k}}$ represent the large-scale
and small-scale fading coefficients, respectively. We further assume
that $g_{mn_{k}},m=1,2,\ldots,M,n=1,2,\ldots,N,k=1,2,\ldots,K_{n},$
are independent identically distributed (i.i.d) $\mathcal{CN}(0,1)$
random variables. In this paper, we also assume that no downlink pilots
will be sent from APs to users. Thus, the transmission in one coherence
time, denoted by $\tau_{c}$ symbols, only includes the uplink training
phase and data multicasting phase which are described in the following
subsections.

\subsubsection{Uplink Training}

Since the TDD mode is adopted and the transmission takes place within
one coherence interval, the channel can be considered reciprocal,
i.e., the channel gains on the uplink and on the downlink are deem
to be identical. Consequently, the APs can estimate the downlink channel
based on the pilot sequences sent by all users on the uplink. Let
$\sqrt{\tau_{p}}\boldsymbol{\psi}_{n}\in\mathbb{C}^{T_{p}\times1}$,
where $\boldsymbol{\psi}_{n}$ be the \emph{common pilot sequence}
transmitted from all users in the $n$-th group, where $\tau_{p}$
is the length of the pilot sequences in symbols. These pilot sequences
are assumed to be independent and orthonormal (i.e., $\Vert\boldsymbol{\psi}_{n}\Vert^{2}=1,\forall n$
and $\boldsymbol{\psi}_{n}\herm\boldsymbol{\psi}_{n^{\prime}}=0,n^{\prime}\neq n$
) among $N$ groups and thus, the effect of pilot contamination is
ignored. The received signal at the $m$-th AP is given by 
\begin{equation}
\mathbf{y}_{m}=\sqrt{\rho_{p}\tau_{p}}\sum_{n=1}^{N}\sum_{k=1}^{K_{n}}h_{mn_{k}}\boldsymbol{\psi}_{n}+\mathbf{w}_{m},
\end{equation}
where $\rho_{p}$ is the power of each pilot symbol, and $\mathbf{w}_{m}\sim\mathcal{CN}(\mathbf{0},\sigma_{n}^{2}\mathbf{I}_{T_{p}})$
is the noise and $\sigma_{n}^{2}$ is the variance of the noise sample.
The $m$-th AP needs to estimate the channel $h_{mn_{k}}$, based
on the received pilot signal $\mathbf{y}_{m}$. In fact, this process
is described in \cite{Doan2017} and is re-derived here with further
details to be self-contained. Specifically, the $m$-th AP projects
$\mathbf{y}_{m}$ onto $\boldsymbol{\psi}_{n}$, producing 
\begin{equation}
\tilde{y}_{mn}=\boldsymbol{\psi}_{n}\herm\mathbf{y}_{m}=\sqrt{\rho_{p}\tau_{p}}\sum_{k=1}^{K_{n}}h_{mn_{k}}+\tilde{w}_{mn},
\end{equation}
where $\tilde{w}_{mn}\triangleq\boldsymbol{\psi}_{n}\herm\mathbf{w}_{m}\sim\mathcal{CN}(0,\sigma_{n}^{2})$.
The minimum mean-square error (MMSE) of the channel estimate is calculated
as
\begin{equation}
\begin{aligned}\hat{h}_{mn_{k}} & =\frac{\mathbb{E}\{h_{mn_{k}}\tilde{y}_{mn}\}}{\mathbb{E}\{\tilde{y}_{mn}^{2}\}}\tilde{y}_{mn}=\frac{\sqrt{\rho_{p}\tau_{p}}\zeta_{mn_{k}}}{\rho_{p}\tau_{p}\sum_{l=1}^{K_{n}}\zeta_{mn_{l}}+\sigma_{n}^{2}}\tilde{y}_{mn}.\end{aligned}
\label{eq:channel_estimate}
\end{equation}
Note that the expectations in the above equation are carried out with
respect to small-scale fading. Since the elements of $\hat{h}_{mn_{k}}$
are independent and identical Gaussian distribution, we can write
it as $\hat{h}_{mn_{k}}=\gamma_{mn_{k}}^{1/2}z_{mn},$ where $\gamma_{mk}=\frac{\rho_{p}\tau_{p}\zeta_{mn_{k}}^{2}}{\sigma_{n}^{2}+\rho_{p}\tau_{p}\sum_{l=1}^{K_{n}}\zeta_{mn_{l}}}$
and $z_{mn}=\frac{\tilde{y}_{mn}}{\sqrt{\rho_{p}\tau_{p}\sum_{l=1}^{K_{n}}\zeta_{mn_{l}}+\sigma_{n}^{2}}}\sim\mathcal{CN}(0,1).$

\subsubsection{Downlink Multicasting}

For the downlink multicasting phase, the APs use the channel estimates
obtained in \eqref{eq:channel_estimate} to form separate radio beams
to the $N$ groups. Similar to \cite{Doan2017}, we adopt \emph{normalized
conjugate beamforming} under the STPC. More specifically, we denote
the symbol to be sent to the $n$-th group by $s_{n}$ such that $\mathbb{E}\big\{|s_{n}|^{2}\big\}=1$.
Then the transmitted symbol from the $m$-th AP is given by 
\begin{equation}
x_{m}=\sqrt{\rho_{d}}\sum\nolimits _{n=1}^{N}\sqrt{\eta_{mn}}\frac{z_{mn}^{\ast}}{\bigl|z_{mn}\bigr|}s_{n},
\end{equation}
where $\eta_{mn}$ is the power control coefficient between the $m$-th
AP and the $n$-th group and $\rho_{d}$ is the maximum power at each
AP. Note that the factor $\frac{z_{mn}^{\ast}}{\bigl|z_{mn}\bigr|}$
in the above is known as normalized conjugate beamforming which incorporates
STPC. Explicitly, the total power constraint at each AP is 
\begin{equation}
\mathbb{E}\big\{\bigl|x_{m}\bigr|^{2}\big\}=\rho_{d}\sum\nolimits _{n=1}^{N}\eta_{mn},\label{eq:powerconst}
\end{equation}
which is \emph{independent} of the small-scale fading coefficient.
We remark that power control is not considered in \cite{Doan2017}.
Finally, the received signal at the $k$-th user in the $n$-th group
is written as 
\begin{equation}
\begin{aligned}r_{n_{k}} & =\sum_{m=1}^{M}h_{mn_{k}}x_{m}+w_{n_{k}}\\
 & =\sqrt{\rho_{d}}a_{n_{k}}s_{n}+\sqrt{\rho_{d}}\sum_{n^{\prime}\neq n}^{N}a_{n_{k}^{\prime}}s_{n^{\prime}}+w_{n_{k}},
\end{aligned}
\label{eq:receivedsig}
\end{equation}
where $a_{n_{k}}=\sum_{m=1}^{M}h_{mn_{k}}\sqrt{\eta_{mn}}\frac{z_{mn}^{\ast}}{\bigl|z_{mn}\bigr|}$,
$a_{n_{k}^{\prime}}=\sum_{m=1}^{M}h_{mn_{k}}\sqrt{\eta_{mn^{\prime}}}\frac{z_{mn^{\prime}}^{\ast}}{\bigl|z_{mn^{\prime}}\bigr|}$,
and $w_{n_{k}}\sim\mathcal{CN}(0,\sigma_{n}^{2})$ is the additive
thermal noise.

\subsubsection{Signal Detection based on Channel Statistics and Spectral Efficiency}

The $k$-th user in group $n$ will rely on the mean of the effective
channel gain to detect $s_{n}$. To see this we rewrite \eqref{eq:receivedsig}
as
\begin{equation}
\begin{aligned}r_{n_{k}} & =\sqrt{\rho_{d}}\ \mathbb{E}\{a_{n_{k}}\}s_{n}+\sqrt{\rho_{d}}\ \big(a_{n_{k}}-\mathbb{E}\{a_{n_{k}}\}\big)s_{n}\\
 & \quad\quad+\sqrt{\rho_{d}}\sum_{n^{\prime}\neq n}^{N}a_{n_{k}^{\prime}}s_{n^{\prime}}+w_{n_{k}}.
\end{aligned}
\end{equation}
As in \cite{Doan2017}, we use the worst-case Gaussian noise argument
given in \cite[section 2.3.4]{marzetta_larsson_yang_ngo_2016} to
obtain the achievable rate (nat/s/Hz) which is expressed as
\begin{equation}
\mathcal{R}_{n_{k}}=\ln\biggl(1+\frac{\rho_{d}\big|\mathbb{E}\{a_{n_{k}}\}|^{2}}{\rho_{d}\Var\{a_{n_{k}}\}+\rho_{d}\sum_{n'\neq n}^{N}\big|\mathbb{E}\{a_{n'_{k}}\}\big|^{2}+\sigma_{n}^{2}}\biggr).\label{eq:AR:gen}
\end{equation}

\begin{prop}
For a multigroup multicast scenario using the normalized conjugate
beamforming, the achievable rate for user $k$ in group $n$ in \eqref{eq:AR:gen}
is reduced to
\begin{equation}
\mathcal{R}_{n_{k}}=\ln\biggl(1+\frac{\frac{\pi\rho_{d}}{4}\big(\sum_{m=1}^{M}\sqrt{\eta_{mn}\gamma_{mn_{k}}}\big)^{2}}{\rho_{d}\sum_{m=1}^{M}\eta_{mn}(N\zeta_{mn_{k}}-\frac{\pi}{4}\gamma_{mn_{k}})+\sigma_{n}^{2}}\biggr).\label{eq:rate}
\end{equation}
\end{prop}
\begin{IEEEproof}
The proof follows the same arguments as those in \cite[Appendix A]{Doan2017},
and, thus, is omitted here due to the space limitation. We remark
that when the power control coefficients are $\eta_{mn}=\frac{1}{N}$,
i.e., EPA, the achievable rate in \eqref{eq:rate} becomes
\begin{equation}
\mathcal{R}_{n_{k}}=\ln\biggl(1+\frac{\frac{\pi\rho_{d}}{4N}\big(\sum_{m=1}^{M}\sqrt{\gamma_{mn_{k}}}\big)^{2}}{\rho_{d}\sum_{m=1}^{M}\bigl(\zeta_{mn_{k}}-\frac{\pi}{4N}\gamma_{mn_{k}}\bigr)+\sigma_{n}^{2}}\biggr),
\end{equation}
which is in fact \cite[Eq. (14)]{Doan2017}.
\end{IEEEproof}

\subsection{Max-min Fairness Power Control}

To ensure the fairness among all the users, we consider the problem
of max-min fairness. Inspired from \cite{Tran:2019:FOMmassiveMIMO,Farooq2020APG},
for the purpose of developing an efficient numerical method, we define
$\mu_{mn}=\sqrt{\eta_{mn}},\forall m,\forall n$. As a result, the
achievable rate in \eqref{eq:rate} is equivalently rewritten as 
\begin{equation}
\mathcal{R}_{n_{k}}(\boldsymbol{\mu})=\frac{\frac{\pi\rho_{d}}{4}\big(\sum_{m=1}^{M}\mu_{mn}\sqrt{\gamma_{mn_{k}}}\big)^{2}}{\rho_{d}\sum_{m=1}^{M}\mu_{mn}^{2}\big(N\zeta_{mn_{k}}-\frac{\pi}{4}\gamma_{mn_{k}}\big)+\sigma_{n}^{2}},
\end{equation}
where $\boldsymbol{\mu}\triangleq[\boldsymbol{\mu}_{1};\boldsymbol{\mu}_{2};\ldots;\boldsymbol{\mu}_{N}]\in\mathbb{R}^{MN}$,
and $\boldsymbol{\mu}_{n}\triangleq[\mu_{1n};\mu_{2n};\ldots;\mu_{Mn}]\in\mathbb{R}^{M},\forall n$.
To ensure that the total transmit power at each AP does not exceed
$\rho_{d}$, we impose the constrain $\sum_{n=1}^{N}\eta_{mn}\leq1$,
which is equivalent to $\sum_{n=1}^{N}\mu_{mn}^{2}\leq1,\forall m$.
The considered power control problem can be mathematically stated
as

\begin{equation}
\begin{aligned}\underset{\boldsymbol{\mu}}{\maximize} & \quad f(\boldsymbol{\mu})=\underset{\forall n_{k}}{\min}\ \mathcal{R}_{n_{k}}(\boldsymbol{\mu})\\
\st & \quad\sum\nolimits _{n=1}^{N}\mu_{mn}^{2}\leq1,\forall m\\
 & \quad\mu_{mn}\geq0,\forall m,\forall n.
\end{aligned}
\tag{\ensuremath{\mathcal{P}}}\label{eq:maxrate:SPC}
\end{equation}

\section{Proposed Solution}

In this section, we propose a low-complexity method for solving \eqref{eq:maxrate:SPC}.
Before doing this, we note that $\mathcal{R}_{n_{k}}(\boldsymbol{\mu})$
is in fact quasi-concave and thus, a bisection method can be applied
to solve \eqref{eq:maxrate:SPC}. To see that, we first rewrite \eqref{eq:maxrate:SPC}
as 
\begin{equation}
\begin{aligned}\underset{\boldsymbol{\mu}}{\maximize} & \quad t\\
\st & \quad\sum\nolimits _{n=1}^{N}\mu_{mn}^{2}\leq1,\forall m\\
 & \quad\mathcal{R}_{n_{k}}(\boldsymbol{\mu})\geq t,\forall n_{k}\\
 & \quad\mu_{mn}\geq0,\forall m,\forall n.
\end{aligned}
\label{eq:maxrate:SPC:epi}
\end{equation}
It is easy to see that the constraint $\mathcal{R}_{n_{k}}(\boldsymbol{\mu})\geq t$
is equivalent to
\begin{multline}
\sqrt{\frac{\pi\rho_{d}}{4}}\big(\sum\nolimits _{m=1}^{M}\mu_{mn}\sqrt{\gamma_{mn_{k}}}\big)\geq\\
\sqrt{e^{t}-1}\sqrt{\rho_{d}\sum\nolimits _{m=1}^{M}\mu_{mn}^{2}\big(N\zeta_{mn_{k}}-\frac{\pi}{4}\gamma_{mn_{k}}\big)+\sigma_{n}^{2}}.
\end{multline}
For a given $t$, the above constraint is indeed a second order cone
constraint. Hence, a bisection search over $t$ can be used to find
the optimal solution. However, the problem with such a method is that
it has high computational complexity which makes it less appealing
to large-scale problems. In what follows, we present solutions to
\eqref{eq:maxrate:SPC} using an APG method introduced in \cite{Li2015}.
For efficiently description of the proposed method, we first reformulate
the problem in the form of a single vector of power control coefficients
as described next.

\subsection{Smoothing Technique}

Let us denote $\bar{\boldsymbol{\mu}}_{m}=[\mu_{m1};\mu_{m2};\ldots;\mu_{mN}]\in\mathbb{R}^{N}$
which include all power control coefficients associated with the $m$-th
AP. The the feasible set in \eqref{eq:maxrate:SPC} can be expressed
as 
\begin{equation}
\mathcal{S}=\{\boldsymbol{\mu}|\boldsymbol{\mu}\geq0;\Vert\bar{\boldsymbol{\mu}}_{m}\Vert^{2}\leq1,\forall m\}.
\end{equation}
Also, to simply the mathematical presentation, we first rewrite $\mathcal{R}_{n_{k}}(\boldsymbol{\mu})$
in a more compact form of $\boldsymbol{\mu}$ as
\begin{equation}
\mathcal{R}_{n_{k}}(\boldsymbol{\mu})=\ln\biggl(1+\frac{\frac{\pi\rho_{d}}{4}(\boldsymbol{\gamma}_{n_{k}}\trans\boldsymbol{\mu}_{n})^{2}}{\rho_{d}\Vert\mathbf{A}_{n_{k}}\boldsymbol{\mu}_{n}\Vert^{2}+1}\biggr),\label{eq:rate2}
\end{equation}
where $\mathbf{A}_{n_{k}}$ is the diagonal defined as
\begin{equation}
\begin{aligned}\mathbf{A}_{n_{k}} & =\diag\Big(\big[\sqrt{N\zeta_{1n_{k}}-\frac{\pi}{4}\gamma_{1n_{k}}};\sqrt{N\zeta_{2n_{k}}-\frac{\pi}{4}\gamma_{2n_{k}}};\\
 & \quad\ldots;\sqrt{N\zeta_{Mn_{k}}-\frac{\pi}{4}\gamma_{Mn_{k}}}\big]\Big).
\end{aligned}
\label{eq:A_nk}
\end{equation}
It is important to note that the objective $f(\boldsymbol{\mu})$
\eqref{eq:maxrate:SPC} is \emph{nonsmooth,} which is a preliminary
requirement for an application of a gradient-based method. To overcome
this issue, we use the smoothing technique introduced in \cite{Nesterov2005a}.
Specifically, for a given \emph{smoothness} parameter $\sigma>0$,
$f(\boldsymbol{\mu})$ is approximated by the following log-sum-exp
function \cite{Nesterov2005a} 
\begin{equation}
f_{\sigma}(\boldsymbol{\mu})=-\frac{1}{\sigma}\ln\Bigl(\frac{1}{NK_{n}}\sum\nolimits _{n=1}^{N}\sum\nolimits _{k=1}^{K_{n}}\exp\bigl(-\sigma\mathcal{R}_{n_{k}}(\boldsymbol{\mu})\Bigr).\label{eq:smoothapprox-1}
\end{equation}
In \cite{Nesterov2005a}, Nesterov proved that $f_{\sigma}(\boldsymbol{\mu})$
is a differentiable approximation of $f(\boldsymbol{\mu})$ with a
numerical accuracy of $\frac{\ln(NK_{n})}{\tau}$, i.e., $f(\boldsymbol{\mu})\leq f_{\sigma}(\boldsymbol{\mu})\leq f(\boldsymbol{\mu})+\frac{\ln(NK_{n})}{\sigma}$.
Hence, with a sufficiently large value of $\sigma$, $f(\boldsymbol{\mu})$
can be replaced with $f_{\sigma}(\boldsymbol{\mu})$ for the optimization
purpose. In this way, \eqref{eq:maxrate:SPC} is approximated by 
\begin{equation}
\begin{aligned}\underset{\boldsymbol{\mu}}{\maximize} & \quad f_{\sigma}(\boldsymbol{\mu})\\
\st & \quad\boldsymbol{\mu}\in\mathcal{S}.
\end{aligned}
\tag{\ensuremath{\hat{\mathcal{P}}}}\label{eq:maxrate:SPC:app}
\end{equation}
In addition to the smoothness of $f_{\sigma}(\boldsymbol{\mu}),$
the projection onto $\mathcal{S}$ can be done in closed form as shall
be seen shortly. This motivates us to apply the APG method in \cite{Li2015}
to solve \eqref{eq:maxrate:SPC:app}.

\subsection{Proposed Accelerated Projected Gradient Method}

The proposed algorithm for solving \eqref{eq:maxrate:SPC:app} is
outlined in Algorithm \ref{alg:APG}.
\begin{algorithm}[tbh]
\caption{Proposed APG Algorithm \label{alg:APG}}

\SetAlgoNoLine
\DontPrintSemicolon
\LinesNumbered 

\KwIn{ $\mathbf{z}^{1}=\boldsymbol{\mu}^{1}=\boldsymbol{\mu}^{0}>0$,
$\sigma>>1$, $\delta>0$, $\alpha_{y}^{0}>0$, $\alpha_{\mu}^{0}>0$,
$0<\kappa<1$}

\For{$n=1,2,\cdots$ }{

Find extrapolated point $\mathbf{y}^{n}$, where $\mathbf{y}^{n}=\boldsymbol{\mu}^{n}+\frac{t_{n-1}}{t_{n}}(\mathbf{z}^{n}-\boldsymbol{\mu}^{n})+\frac{t_{n-1}-1}{t_{n}}(\boldsymbol{\mu}^{n}-\boldsymbol{\mu}^{n-1})$.

Find the smallest nonnegative integer $l_{y}$ and $\mathbf{z}^{n+1}$
such that $f_{\sigma}\bigl(\mathbf{z}^{n+1}\bigr)\geq f_{\sigma}\bigl(\mathbf{y}^{n}\bigr)+\delta\bigl\Vert\mathbf{z}^{n+1}-\mathbf{y}^{n}\bigr\Vert^{2}$,
where $\mathbf{z}^{n+1}=P_{\mathcal{S}}\big(\mathbf{y}^{n}+\kappa^{l_{y}}\alpha_{y}^{n-1}\nabla f_{\sigma}(\mathbf{y}^{n})\big)$.\\

Find the smallest nonnegative integer $l_{\mu}$ and $\mathbf{v}^{n+1}$
such that $f_{\sigma}\bigl(\mathbf{v}^{n+1}\bigr)\geq f_{\sigma}\bigl(\boldsymbol{\mu}^{n}\bigr)+\delta\bigl\Vert\mathbf{v}^{n+1}-\boldsymbol{\mu}^{n}\bigr\Vert^{2}$,
where $\mathbf{v}^{n+1}=P_{\mathcal{S}}\big(\boldsymbol{\mu}^{n}+\kappa^{l_{\mu}}\alpha_{\mu}^{n-1}\nabla f_{\sigma}(\boldsymbol{\mu}^{n})\big)$.\\

Set $\alpha_{y}^{n}=\kappa^{l_{y}}\alpha_{y}^{n-1}$, $\alpha_{\mu}^{n}=\kappa^{l_{\mu}}\alpha_{\mu}^{n-1}$,
and extrapolation parameter $t_{n+1}\triangleq0.5+\sqrt{t_{n}^{2}+0.25}$.\\

\eIf{$f_{\sigma}\bigl(\mathbf{z}^{n+1}\bigr)>f_{\sigma}\bigl(\mathbf{v}^{n+1}\bigr)$
} {

$\boldsymbol{\mu}^{n+1}=\mathbf{z}^{n+1}$ } {

$\boldsymbol{\mu}^{n+1}=\mathbf{v}^{n+1}$

}

}
\end{algorithm}
From \eqref{eq:smoothapprox-1}, the gradient of $f_{\sigma}(\boldsymbol{\mu})$
is found as 
\begin{equation}
\frac{\partial}{\partial\boldsymbol{\mu}}f_{\sigma}(\boldsymbol{\mu})=\frac{\sum\nolimits _{n=1}^{N}\sum_{k=1}^{K_{n}}\Bigl(\exp\bigl(-\sigma\mathcal{R}_{n_{k}}(\boldsymbol{\mu})\bigr)\nabla_{\boldsymbol{\mu}}\mathcal{R}_{n_{k}}(\boldsymbol{\mu})\Bigr)}{\sum\nolimits _{n=1}^{N}\sum_{k=1}^{K_{n}}\exp\bigl(-\sigma\mathcal{R}_{n_{k}}(\boldsymbol{\mu})\bigr)}.\label{eq:gradf}
\end{equation}
It is easy to see that the gradient of $\mathcal{R}_{n_{k}}(\boldsymbol{\mu})$
is 
\begin{equation}
\nabla_{\boldsymbol{\mu}}\mathcal{R}_{n_{k}}(\boldsymbol{\mu})=\frac{\nabla_{\boldsymbol{\mu}}\big(b_{n_{k}}(\boldsymbol{\mu}_{n})+c_{n_{k}}(\boldsymbol{\mu}_{n})\big)}{b_{n_{k}}(\boldsymbol{\mu}_{n})+c_{n_{k}}(\boldsymbol{\mu}_{n})}-\frac{\nabla_{\boldsymbol{\mu}}c_{n_{k}}(\boldsymbol{\mu}_{n})}{c_{n_{k}}(\boldsymbol{\mu}_{n})},
\end{equation}
where $b_{n_{k}}(\boldsymbol{\mu}_{n})\triangleq\frac{\pi\rho_{d}}{4}(\boldsymbol{\gamma}_{n_{k}}\trans\boldsymbol{\mu}_{n})^{2}$
and $c_{n_{k}}(\boldsymbol{\mu}_{n})\triangleq\rho_{d}\Vert\mathbf{A}_{n_{k}}\boldsymbol{\mu}_{n}\Vert^{2}+1.$
By recalling the identity $\nabla_{\mathbf{x}}\Vert\mathbf{A}\mathbf{x}\Vert^{2}=2\mathbf{A}\trans\mathbf{A}\mathbf{x}$
for any symmetric matrix $\mathbf{A}$, the gradients $\nabla_{\boldsymbol{\mu}}b_{n_{k}}(\boldsymbol{\mu}_{n})$
and $\nabla_{\boldsymbol{\mu}}c_{n_{k}}(\boldsymbol{\mu}_{n})$ in
the above equation are calculated as 
\begin{equation}
\nabla_{\boldsymbol{\mu}}b_{n_{k}}(\boldsymbol{\mu}_{n})=\frac{\pi\rho_{d}}{2}\boldsymbol{\gamma}_{n_{k}}\boldsymbol{\gamma}_{n_{k}}\trans\boldsymbol{\mu}_{n},\label{eq:BkGradient}
\end{equation}

\begin{equation}
\nabla_{\boldsymbol{\mu}}c_{k}(\boldsymbol{\mu})=2\rho_{d}\mathbf{A}_{n_{k}}\trans\mathbf{A}_{n_{k}}\boldsymbol{\mu}_{n}.\label{eq:CkGradient}
\end{equation}

Further note that the projection of any vector $\mathbf{x}$ onto
the $\mathcal{S}$ is defined as
\begin{equation}
P_{\mathcal{S}}(\mathbf{x})=\arg\min\bigl\{||\mathbf{x}-\mathbf{u}||\ |\ \mathbf{u}\in\mathcal{S}\bigr\}.\label{eq:projection}
\end{equation}
The Euclidean projection onto $\mathcal{S}$ defined in \eqref{eq:projection}
can be done can be done \emph{in parallel} and by \emph{closed-form
expressions}. In particular, the optimization problem in \eqref{eq:projection}
can be decomposed into sub-problems at each AP $m$ as
\begin{equation}
\bar{\boldsymbol{\mu}}_{m}=\arg\min\bigl\{||\bar{\mathbf{x}}_{m}-\bar{\boldsymbol{\mu}}_{m}||\ |\ \Vert\bar{\boldsymbol{\mu}}_{m}\Vert^{2}\leq1,\boldsymbol{\bar{\mu}}_{m}\geq0\bigr\},\label{eq:projection-1}
\end{equation}
where $\bar{\mathbf{x}}_{m}=[x_{m1};x_{m2};\ldots;x_{mN}]\in\mathbb{R}^{N}.$
The above problem can solved by finding the projection onto the intersection
of the positive orthant and Euclidean ball \cite[Theorem 7.1]{Bauschke2017}.
More specifically, we first project $\bar{\mathbf{x}}_{m}$ onto the
positive orthant, i.e., $[\bar{\mathbf{x}}_{m}]_{+}$ and then onto
the unit-norm ball which is simply given by
\begin{equation}
\bar{\boldsymbol{\mu}}_{m}=\begin{cases}
[\bar{\mathbf{x}}_{m}]_{+} & \Vert[\bar{\mathbf{x}}_{m}]_{+}\Vert\leq1,\\
\frac{[\bar{\mathbf{x}}_{m}]_{+}}{\Vert[\bar{\mathbf{x}}_{m}]_{+}\Vert} & \textrm{otherwise}.
\end{cases}\label{eq:projEuclidean}
\end{equation}

\subsection{Complexity Analysis}

Now, we describe the complexity of the proposed algorithm using the
big-O notation. Note that for each general step in Algorithm \ref{alg:APG},
three factors contribute towards the computational complexity; the
objective \eqref{eq:smoothapprox-1}, the gradient \eqref{eq:gradf}
and the projection \eqref{eq:projEuclidean}. It can be easily verified
that the computation of $\mathcal{R}_{n_{k}}$ requires $M$ multiplications
and therefore, the complexity of finding the objective is $\mathcal{O}\bigl(M\sum_{n=1}^{N}K_{n}\bigr)$.
Similarly the gradient $\frac{\partial}{\partial\boldsymbol{\mu}}f_{\sigma}(\boldsymbol{\mu})$
has the complexity of $\mathcal{O}\bigl(M\sum_{n=1}^{N}K_{n}\bigr)$
also. The projection operation requires the computation of $l_{2}$-norm
of $\mathbb{R}^{N}$ vectors at all $M$ APs and thus, has complexity
of $\mathcal{O}(MN)$. In summary, the per-iteration complexity of
the proposed algorithm is $\mathcal{O}\bigl(M\sum_{n=1}^{N}K_{n}\bigr)$.

\section{Numerical Results}

In this section, we evaluate the performance of the proposed method
in different multigroup multicasting cell-free massive MIMO scenarios.
The system bandwidth is set to $B=\SI{20}{\mega\hertz}$ and the carrier
frequency to $f=\SI{1900}{\mega\hertz}$. We generate the channel
in \eqref{eq:channel} similar to \cite{Doan2017}, where $\sigma_{\mathrm{sh}}=9$
dB be the standard deviation of the log-normal shadowing. Also, the
noise power is calculated as $N_{0}=k\times T\times B\times NF$,
where $NF=\SI{9}{\dB}$ is the noise figure, $T=\SI{290}{\kelvin}$
is the temperature, and $k=1.38\times10^{-23}\SI{}{\joule/\kelvin}$
is the Boltzmann's constant. Further, we choose $\rho_{d}=\rho_{p}=\SI{0.2}{\watt}$,
$\tau_{p}=20$, $\tau_{c}=200$ and $K_{n}=K$ (i.e., same number
of users for each group) in all the experiments. APs and users are
distributed uniformly over the area of $D=\SI{1}{\km\squared}$. The
parameters involved in Algorithm \ref{alg:APG} are set to $\sigma=100$,
$\delta=10^{-5}$ and $\kappa=0.45$.

\begin{figure}[tbh]
\centering\includegraphics[viewport=62bp 558bp 296bp 741bp,clip,width=0.75\columnwidth,height=0.55\columnwidth]{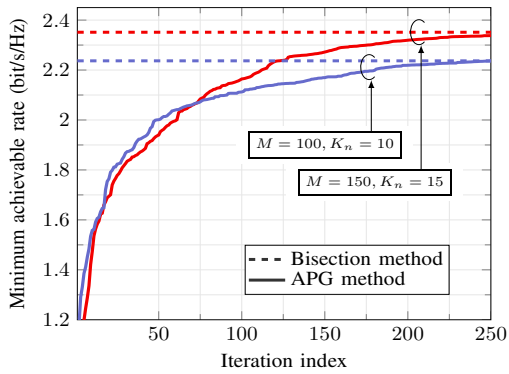}\caption{Comparison of convergence of Algorithm \ref{alg:APG} with the bisection
method for two scenarios; $M=100,K_{n}=10$ and $M=150,K_{n}=15$.
Here, we consider four groups with $K_{n}=K$ users in each group.}
\label{fig:convergence}
\end{figure}

First, we plot in Fig. \ref{fig:convergence} the achieved minimum
rate of all users using Algorithm \ref{alg:APG} for two different
scenarios. Note that one set of channel realizations is randomly generated
for each scenario. In particular, we compare the convergence of the
proposed APG method with the the bisection method. To solve the resulting
feasibility problem in each iteration of the bisection method, we
use the modeling tool CVX \cite{cvx}. 
It can be observed from the Fig. \ref{fig:convergence} that the proposed
method reaches the same objective value as the bisection method. The
advantage of Algorithm \ref{alg:APG} is that it takes much less run
time than the bisection method to return a solution as recorded in
Table \ref{table: Table 1}. Particularly, the bisection method cannot
handle large-scale scenarios due to the large required memory and
extremely long run time.
\begin{table}[tbh]
\caption{Comparison of run-time (in seconds) for $N=2$ and $K_{n}=15$. }
\label{table: Table 1}
\centering{}%
\begin{tabular}{c|c|c}
\hline 
APs & Bisection Method & Proposed APG Method\tabularnewline
\hline 
100 & 54.77  & \textbf{6.43 }\tabularnewline
\hline 
150 & 68.50  & \textbf{13.58 }\tabularnewline
\hline 
200 & 103.75 & \textbf{26.69 }\tabularnewline
\hline 
\end{tabular}
\end{table}

\begin{figure}[tbh]
\centering\includegraphics[viewport=62bp 550bp 289bp 738bp,clip,width=0.75\columnwidth,height=0.55\columnwidth]{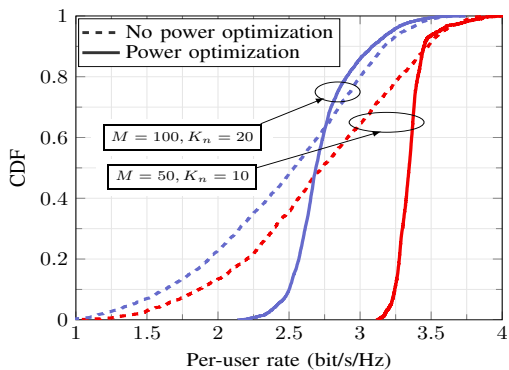}\caption{CDF for the considered power control scheme compared with the EPA
scheme. In the experiment, we take $N=2$ and simulate two scenarios;
$M=50,K=10$ and $M=100,K=20$.}
\label{fig:cdf}
\end{figure}

Next, we demonstrate the benefits of power control optimization for
multigroup multicast cell-free massive MIMO systems. To this end,
we plot in Fig. \ref{fig:cdf} the achieved cumulative distribution
function (CDF) of per-user rate using the proposed power control algorithm
and compare it with the EPA scheme in \cite{Doan2017}. 
The results shown in Fig. \ref{fig:cdf} are interesting in many ways.
First, the considered power control scheme outperform the EPA method
in terms of performance for both considered scenarios. Another observation
is that Algorithm \ref{alg:APG} is better in terms of fairness among
the users. Note that for the fixed area, the per-user rate decreases
as the problem size increases. This is due to the fact that with an
increase in the number of users, the inter-user interference among
the users of the different groups increases which in turn causes a
significant decrease in the achievable rate.

\section{Conclusion}

We have considered the max-min fairness problem in the downlink channel
of multigroup multicasting cell-free massive MIMO. We have formulated
the power control problem using normalized conjugated beamforming
scheme which incorporates the STPC to strictly constrain the downlink
power to stay under maximum allowable power at each AP. To solve the
problem, we have proposed a low-complexity algorithm based on the
APG iterations. Our simulation results have shown that the proposed
algorithm achieves the same objective as the well-known bisection
algorithm but in much lesser run time. More specially, the proposed
APG method outperforms the EPA method both in terms of achievable
rate fairness among the users.

\section*{Acknowledgment}

This publication has emanated from research supported by a Grant from
Science Foundation Ireland under Grant number 17/CDA/4786.

\bibliographystyle{IEEEtran}
\bibliography{IEEEabrv,bibTex,references}

\end{document}